\documentstyle{article}

\begin{document}
\begin{center}
\large\bf
Vortex Drag in Quantum Hall Effect.
\normalsize
\vskip 0.3cm

S.A.Vitkalov
\vskip 0.3cm

\em Chemistry Department, University of Florida,\\
P. N. Lebedev Physical Institute, Russian Academy of Science, Moscow 
\vskip 1cm
\normalsize
Published in Pis'ma v ZheTP 67, 276-280 (1998)
\end{center}

\vskip 2cm
A new model of momentum and electric field transfer between two
adjacent 2D electron systems in the Quantum Hall Effect is proposed. 
The drag effect is due to momentum transfer from the 
vortex system of one layer to the vortex system of another layer.
The remarkable result of this approach is periodical change of $sign$ 
of the dragged electric field with difference between the layer
filling factors.

\pagebreak
\normalsize
Drag effect in double layer two dimensional electron systems  (2DES) has 
been the subject of intense recent interest, especially at high magnetic 
fields, where the Quantum Hall Effect (QHE) exists [1].
Theoretical explanations of the observed drag effect are mostly reduced to
the ordinary Coulomb interaction between electrons occupying the 
Landau levels in the two layers [2]. 
These explanations are not much related to the Quantum Hall Effect. 
I propose a new mechanism for the momentum transfer
between the layers, existing exclusively under the QHE conditions. 
For the simplicity I consider integer filling factors. 

The model is based on a consideration of the Quantum Hall liquid as a
superfluid state of the Chern-Simons charged bosons [3,4]. The ground state 
$\phi=\phi_1+i\phi_2$
have a quasilong range phase correlation [5] and is uniform in the mean field
approximation at  "magic" filling factors $\nu_i^0=i$, where $i$=1, 2... is
an integer. 
 
In accordance with this point of view on the QHE, a vortex ($H>H_i^0$) or
an antivortex ($H<H_i^0$)
excitations are
created away from the filling factor $\nu_i^0$ with the concentration 
$N^v_i=\vert H-H_i^0\vert /\Phi_0$,
where H is the external magnetic field, $H_i^0$ is a magnetic field 
corresponding to the filling factor $\nu_i^0$. 
The magnetic flux carried by the vortex (antivortex) is   
$\Phi_0$ ($-\Phi_0$), where $\Phi_0=hc/e$ is the flux quantum.
If the ground state carries a supercurrent $\vec j^{ext}$
the vortices experience an average force $\vec F^{ext}$, which 
corresponds to the force acting on the vortices in ordinary
superconductor. Formally the force is a result of current-current
interactions  between the external supercurrent $\vec j^{ext}$ and 
the vortex supercurrent $\vec j^v$ [6]. Hamiltonian of the interaction is  
$H_{int}=\Lambda (\vec j^v \vec j^{ext})$, where $\Lambda$ is an 
interaction constant.
The negative derivative of the Hamiltonian $H_{int}$ with respect to 
the vortex position is the external  force $\vec F^{ext}$. 
The gauge invariant expression for the supercurrent of the charged bosons 
is the same as for ordinary superconductor, 
but the charge of the Cooper pair $2e$
needs to be changed by the boson charge $q$. 
The charged boson supercurrent is  
$\vec j=-(\Lambda c)^{-1}(\vec A-hc/2\pi q\nabla \chi)$, where $\chi$ is
the  phase of the boson ground state $\phi$.
Considering the function $\phi$ as one valued function, 
the external force $F^{ext}$ is found to be: 

$$\vec F^{ext}=\frac {\pm \Phi_0}{q_ic}[\vec j^{ext}\times\vec e_z] 
\eqno{(1)}$$

where $c$ is the light velocity, $\vec e_z$ is direction of 
the magnetic field $\vec H$, 
and the sign $+$ ($-$) corresponds to the vortex (antivortex).  
The value $q_i$ is
the boson charge in the electron charge units : $q=q_ie$.
A similar expression for the force in an external electric field $\vec E$
was found in a different approach [7]:

$$ \vec F_2^{ext}=\frac {e^2}{hc}\Phi_0 \vec E \eqno{(1a)}$$

Using the relation between the current and the electric field in the QHE : 
$j^{ext}=\sigma_{xy} E=q_i\times e^2/hE$, we obtain $F^{ext}=F^{ext}_2$.

There are several possibilities of the vortex motion in the external current
$\vec j^{ext}$. 
A simple picture 
is considered here. 
At the temperature $T=0K$ all vortices are pinned by the disorder 
and the current $\vec j^{ext}$
flows without any dissipation. There is no momentum transfer into the 
vortex system at $T=0K$. 
At a finite temperature $T>0K$, the vortices can jump from one
point to another one due to the thermal fluctuations. The external current
$\vec j^{ext}$ induces an average momentum (creep) of the vortices 
in the direction of the force $\vec F^{ext}$. 
Thus, at the temperature $T>0K$ there is a momentum transfer from the external
current to the vortex system. The average nonzero momentum of the vortices
in the first layer will relax, 
and partially will transform into the momentum of vortices of
the second layer. 
Actually, the most effective channel of the momentum transfer
between the layers is not clear. A suitable candidate is the phonon
system. 

In this paper the momentum transfer between the layers is studied 
phenomenologically. 
Let us consider the momentum transfer between the vortices at some QHE
resistance minimum
of the first layer ($H_1^0$) and the vortices at  some QHE resistance 
minimum of
the second layer($H_2^0$) (see fig.1).  
The average momentum of the vortices in the first (second) 
layer is $\vec P_1$ ($\vec P_2$).
The total force from the external current $\vec j^{ext}$, acting on the 
vortex system, 
(layers are squares with unit areas) is 
$N_i^v \vec F^{ext}_i$ (1), 
where $i$=1,2 is the layer index now.  

Newton's equations for the vortices momentum are ($H-H_i^0=N_i^v\Phi_0$):

$$ d\vec P_1/dt=1/(q_1c)[\vec j_1^{ext}\times (\vec H-\vec H_1^0)] 
- \vec P_1/\tau_1 - \vec F^{int} \eqno{(2)}$$ 

$$ d\vec P_2/dt=1/(q_2c)[\vec j_2^{ext}\times(\vec H-\vec H_2^0)] 
- \vec P_2/\tau_2 + \vec F^{int} \eqno{(3)}$$

where $\tau_i$ are the momentum relaxation rates and $\vec F^{int}$ is the
interlayer drag force. 

In the experiment [1], the additional electric field $\vec E^{ext}_2$ is
applied to the second layer to cancel the current $\vec j^{ext}_2$. 
In this case the equation (3) is transformed into:

$$ d\vec P_2/dt= - \vec P_2/\tau_2 + \vec F^{int} \eqno{(3a)}  $$

At a small perturbation of the vortex distribution function 
the interlayer drag
force is proportional to the vortex momentum $\vec P_1$. 
At a small value of vortex concentration $N_2^v$, the
total drag force is proportional to the concentration $N_2^v$. 
Therefore the interlayer drag force $\vec F^{int}$ 
can be approximated by: 

$$\vec F^{int}=\alpha N_2^v \vec P_1  \eqno{(4)}$$

where $\alpha$ is a constant.

Since the interlayer drag force $F^{int}$ is much less than the external force
$F^{ext}$, the first one can be taken as a small perturbation in
the eq.(2). 
From the eq.(2), (3a) and (4) the dragged momentum of the vortices 
$\vec P_2$ is found to be:

$$\vec P_2=\frac{\alpha\tau_1\tau_2}{q_1c}[\vec j^{ext}_1\times
(\vec H-\vec H^0_1)]N_2^v \eqno{(5)}$$

In accordance with the Maxwell electrodynamics
electric field generated by the vortex movement is 
$\vec E_2=1/c[(\vec H-\vec H_2^v) \times \vec V_2$], where $\vec V_2$ is
the 
average vortex velocity in the second layer.  The external electric
field $\vec E_2^{ext}$ should cancel the electric field $\vec E_2$ 
to prevent the total current
$j^{ext}_2$:  $\vec E_2+\vec E_2^{ext}=0$.  

Using the expression $\vec P_2=m_vN_2^v \vec V_2$, 
where $m_v$ is a mass of the vortex, 
the dragged electric field in the second layer is found to be:

$$\vec E_2=g\vec j_1^{ext}(H-H_1^0)(H-H_2^0) \eqno{(6)}$$

where $g=\alpha\tau_1\tau_2/(q_1m_vc^2)$ is a constant.

1)
Let's consider a case of the 
filling factors $\nu_1=\nu_2+i$ (fig.1a), where $i$ is an integer
(for example, equal electron concentrations $n_1^e=n_2^e$ in both layers).
In this case the electric field in the second layer is 

$$\vec E_2=g\vec j_1^{ext}(H-H_1^0)^2 \eqno{(7)}$$

In eq.(7) the drag voltage has a quadratic dependence on the magnetic field
deviation from the QHE resistance minimum $H_1^0$. 
Far from the minimum  $H_1^0$ the phase
coherent ground state $\phi$ breaks and the vortex drag disappears. 
Thus each
QHE minimum is accompanied by two picks of the drag voltage 
$E_2$ on 
different sides away from the minimum $H_1^0$. This behavior correlates with
the experiment [1].       

2) Let's make the QHE resistance minimums $H_1^0$ and $H_2^0$ in the 
layers not coincide with each other(fig. 1b).  
In this case, in accordance with the eq.(6), the
electric field $E_2$ changes the $sign$ in the interval $H_1^0<H<H_2^0$.
It happens because of the antivortices ($H<H_2^0$) are dragged by the 
vortices ($H>H_1^0)$. 

If we fix the electron concentration in the first
layer $n^e_1$ and vary the concentration in the second layer
$n^e_2$
the dragged voltage sign will $oscillate$,  
due to the periodicity of the QHE
conditions with the electron concentration $n_e$ [8].
The main reason of the sign variations is 
the periodic change of the ground state excitations     
from the vortices to the 
antivortices  with the electron concentration or  with the
external magnetic field.

In the conclusion, the vortex model of the drag effect in the bilayer two
dimensional electron systems is proposed. 
Arising  exclusively under  the Quantum
Hall Effect conditions  
the drag effect is induced by the momentum transfer from the
vortex
excitations of the ground state of the first layer to the vortex excitations 
of the ground
state of the second layer.
At equal electron concentrations in the layers the dragged voltage has
the double picks structure, in accordance with the experiment [1].
For the vortex-vortex or antivortex-antivortex momentum transfer, 
the sign of
the external electric field 
in the second layer is opposite to the sign of the
electric
field in the first layer, in accordance with the experiment [1].
For the vortex-antivortex or antivortex-vortex momentum transfer, 
the sign of
the  external electric 
field in the second layer is the $same$ as the sign of
the electric field in the first layer. 

This work was supported by National Science Foundation under Grant No.
CHE-9624243 and by program "Solid State Nanostructures Physics" 97-1050.

\vskip 1cm
\normalsize

1. H. Rubel, A. Fisher, W. Dietsche, K. von Klitzing and K. Eberl,
Phys. Rev. Lett. 78, 1763, 1997

2. M. C. Bonsager, K. Flensberg, B. Yu-Kuang Hu and Antii-Pekka Janho
Phys. Rev. Lett. 77, 1366, 1996

3. S.-C. Zhang Int. J. Mod. Phys. B 6, 25, 1992

4. S. Kivelson, D.-H. Lee, S.-C. Zhang Phys. Rev. B 46, 2223, 1992

5. S. M. Girvin, A. H. Macdonald Phys. Rev. Lett. 58, 1252, 1987

6. А. А. Абрикосов "Основы теории металлов"  стр.396, Москва, Наука,1985

7. J. E. Avron and P. G. Zograf Preprint cond-mat/9711177, LANL e-print
archive, 1997

8. K. von Klitzing, G. Dorda, M. Pepper Phys. Rev. Lett. 45, 494,(1980) 

\vskip 1cm
Figure

Dependence of the drag resistance $R_{drag}=E_2/j_1^{ext}$ 
on magnetic field H in bilayer 2D electron
system under Quantum Hall Effect conditions. 
Sign of the $R_{drag}$ is positive in the
case of filling factors $\nu_1=\nu_2+i$.
The sign of the $R_{drag}$ is negative in the case of filling factors 
$\nu_1=\nu_2+i+1/2$, where $i$ is an integer and 1,2 are layer indexes.

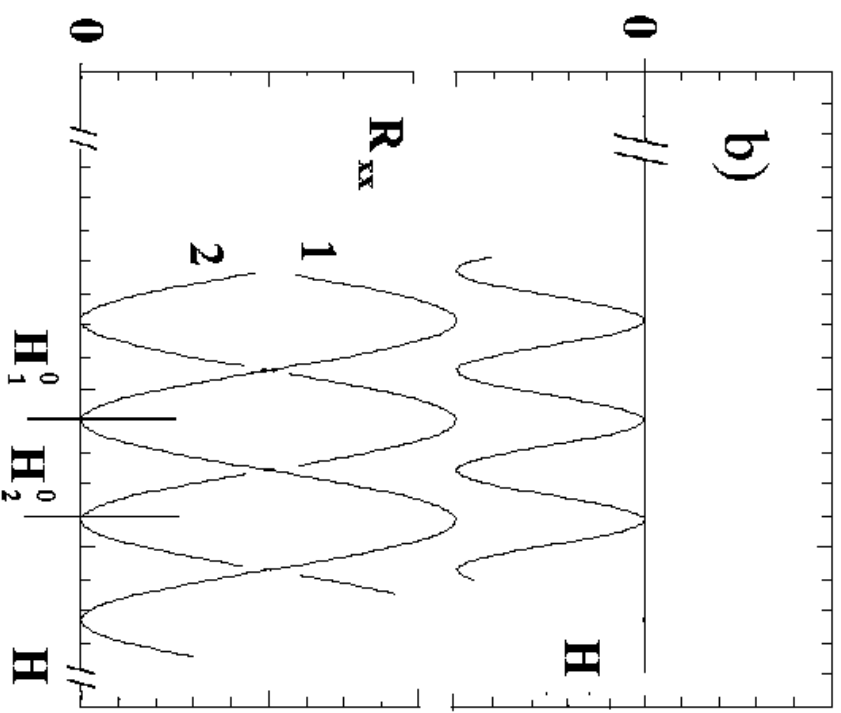
\end{document}